\documentclass[12pt,a4paper]{article}
\font\cst=cmr10 scaled \magstep4
\font\csc=cmr10 scaled \magstep2

\begin{document}

\vglue 1.5cm
\centerline{\cst Smooth branes and junction conditions}
\vskip 0.7cm
\centerline{\cst in Einstein Gauss-Bonnet gravity}
\vskip 1.5cm

\centerline{\bf Nathalie Deruelle}
\vskip 0.5 cm
\centerline{\it  Institut d'Astrophysique de Paris,}
\centerline{\it GReCO, FRE 2435 du CNRS,}
\centerline{\it 98 bis Boulevard  Arago, 75014 Paris, France}

\centerline{and}

\centerline{\it Yukawa Institute for Theoretical Physics}
\centerline{\it Kyoto University, Kyoto 606-8502, Japan}

\vskip 0.7cm
\centerline{\bf Cristiano Germani}
\vskip 0.2cm
\centerline{\it Institute of Cosmology and Gravitation}
\centerline{\it University of Portsmouth}
\centerline{\it Portsmouth PO1 2EG, England}

\centerline{and}

\centerline{\it Universit\`a la Sapienza di Roma}
\centerline{\it Dipartimento di Ingegneria Aerospaziale e Astronautica}
\centerline{\it Via Eudossiana 16, 00184 Roma, Italy}

\medskip
\vskip 0.8cm
\centerline{June 26th 2003}

\vskip 1.5cm

\noindent
{\bf Abstract}
\bigskip

\noindent
Using ``smooth brane" solutions of the field equations, we give an alternative derivation of the junction conditions for a ``brane" in a five dimensional ``bulk", when gravity is
governed by the Einstein Lanczos (Gauss-Bonnet) equations.

\vfill\eject

\noindent
{\csc I. Introduction}
\medskip

Higher dimensional gravity theories based on the Lanczos  Lagrangian or its generalization by Lovelock (also called Gauss-Bonnet or Euler Lagrangians), which are non-linear in
the curvature but such that the field equations remain second order in the metric coefficients, have been known for a long time, see~\cite{DerMad03} for early
references and a recent review.

They have recently attracted renewed interest motivated by the invention of  ``brane scenarios" in which the observable universe is described as a four dimensional
singular surface, or ``brane", of a five dimensional space-time, or ``bulk" obeying Einstein's equations for gravity (see~\cite{RanSun99b} for basic references and a recent
review). 

The extension of these brane models to gravity theories based on the Einstein Gauss-Bonnet Lagrangian (see [3-4]) has however been plagued by the problem of generalizing the Israel junction conditions~\cite{Isr66} which describe gravity on the brane. The reason for the difficulty  is that the
field equations are only quasi-linear in the second order derivatives of the metric coefficients (see e.g.~\cite{DerMad03}). As a result conflicting linearized equations for
brane gravity [3-4] and conflicting cosmological models [6-8] can be found in the recent literature. A physical explanation for these differences has been given in~\cite{Ger03} using thick brane models.

In this contribution, using an adequate definition of the brane stress energy tensor,  we confirm the results obtained in~\cite{Dav02} (which agree with [4] and [6] and generalize the ``total bending" junction conditions of~\cite{Ger03}). To do so we use an approach, directly based on the field
equations rather than on considerations on the proper boundary terms to be added to the action. More precisely we consider  smooth ``brane" solutions of the field equations for
gravity coupled to a confining scalar field and show that they tend, in the limit of infinite thinness, to a solution for a thin brane endowed with matter whose stress energy
tensor is the one given by the general junction conditions obtained in~\cite{Dav02}.

\bigskip
\noindent
{\csc II. The thin brane problem in Einstein Gauss-Bonnet theory~: a summary}
\medskip

To construct a ``$Z_2$-symmetric braneworld" one considers a 5-dimensional manifold $V_+$ with a timelike edge~; one then makes a copy $V_-$ of $V_+$ and superposes the copy and the
original manifold onto each other along the edge (this is the so-called $Z_2$ symmetry)~; one thus obtains a spacetime, or braneworld, composed of a ``bulk" $V_5$, and
 a singular surface, or ``brane" $\Sigma_4$, whose extrinsic curvature is discontinuous : the extrinsic curvature of $\Sigma_4$ embedded in $V_-$ is the opposite of its extrinsic curvature as embedded in $V_+$. 

Suppose now that gravity in the bulk $V_5$ is described by the vacuum Einstein Lanczos (Gauss-Bonnet) equations, that is
$$\sigma^A_{[2]B}\equiv\Lambda \delta^A_B+\sigma^A_B+\alpha\, \sigma^A_{(2)B}=0\eqno(2.1)$$
$\Lambda$ being the bulk ``cosmological constant", $\alpha(>0)$ some $(length)^2$  parameter and the Einstein and Lanczos tensors being defined as
$$\sigma^A_B\equiv r^A_B-{1\over2}\delta^A_B\,s\,,$$
$$\sigma^A_{(2)B}\equiv 2\left[R^{ALMN}R_{BLMN}-2r^{LM}R^A_{\ LBM}-2 r_A^Lr_{LB}+
sr^A_B\right]-{1\over2}\delta^A_B L_{(2)}\,,\eqno(2.2)$$
$$L_{(2)}\equiv s^2-4r^{LM}r_{LM}+R^{LMNP}R_{LMNP}$$
where $R^A_{\ BCD}\equiv\partial_C\Gamma^A_{BD}-...$ are the components of the  Riemann tensor,  $\Gamma^A_{BD}$ being the Christoffel symbols, all indiced being moved with the metric $g_{AB}$ and its inverse $g^{AB}$~; $r_{BD}\equiv R^A_{\
BAD}$ are the Ricci tensor components, $s\equiv g^{BD}r_{BD}$ is the scalar curvature.

Suppose also, for definitiveness,  that the bulk is locally anti-de Sitter spacetime. Then, because of maximal symmetry,
$$R_{ABCD}=-{1\over {\cal L}^2}(g_{AC}g_{BD}-g_{BC}g_{AD})\eqno(2.3)$$
with the characteristic length scale ${\cal L}$ given by
$${1\over {\cal L}^2}={1\over4\alpha}\left(1\pm\sqrt{1+{4\alpha\Lambda\over3}}\right)
\eqno(2.4)$$
in order to satisfy (2.1). 

Finally suppose, for the sake of the argument, that $\Sigma_4$ is flat.
A convenient coordinate system to describe the almost everywhere anti-de Sitter braneworld is, in that case
$$ds^2|_5=dw^2+e^{-2|w|/{\cal L}}\eta_{\mu\nu}dx^\mu dx^\nu\eqno(2.5)$$
with $\eta_{\mu\nu}=(-, +,+,+)$ the Minkowski metric, where $w>0$ spans $V_+$ and $w<0$ spans $V_-$ and where the brane is located at $w=0$ (and ${\cal L}>0$).

The extrinsic curvature of $\Sigma_4$ in $V_5\ {\cal t}\ \Sigma_4$ is discontinuous~:
$$K_{\mu\nu}={1\over{\cal L}}\eta_{\mu\nu}\,{\cal S}(w)\eqno(2.6)$$
where the sign distribution  ${\cal S}(w)$ is +1 if $w>0$, and -1 if $w<0$. Some components of the braneworld Riemann tensor therefore exhibit a delta-type singularity (since
${\cal S}'(w)=2\delta(w)$) and one expects that the braneworld $V_5\ {\cal t}\ \Sigma_4$ satisfies the Einstein Lanczos (Gauss-Bonnet) equations everywhere---that is, $\Sigma_4$ included--- but in the
presence of ``matter" localised on
$\Sigma_4$, i.e. that one has, in $V_5\ {\cal t}\ \Sigma_4$ :
$$\sigma^A_{[2]B}= T^A_B\,{\cal D}(w)\eqno(2.7)$$
where ${\cal D}$ is a distribution localized on $\Sigma_4$, i.e. proportional to some linear
combination of the Dirac delta distribution and its derivatives and
where $T^A_B$ is interpreted as the stress-energy tensor of ``tension plus matter" in the brane.
If ${\cal D}$ is the Dirac distribution (and we shall see that this is indeed the case) then (2.7) relates the ``total bending" of the brane to its stress-energy tensor, as~\cite{Ger03}
$$\int_{-\infty}^{+\infty}\sigma_{[2]B}^A\, dw=T^A_B\,.$$

The question now is to express this ``stress-energy" tensor in terms of the discontinuity of the extrinsic curvature.

In the simple example of a flat brane in AdS$_5$ it is a straightforward exercise to find that $\sigma^\mu_{[2]\nu}$ (and only $\sigma^\mu_{[2]\nu}$) possesses a part which is confined
on
$\Sigma_4$~:
$$\sigma^\mu_{[2]\nu}=-6\,\delta^\mu_\nu\,
{\delta(w)\over{\cal L}}\left[1-4{{\cal S}^2(w)\over{\cal L}^2}\right]\,.\eqno(2.8)$$

Hence, in pure Einstein theory ($\alpha=0$, ${6\over{\cal L}^2}=-\Lambda$) one recovers the well-known result~\cite{RanSun99b}
$$T^w_w=T^w_\mu=0\quad,\quad T^\mu_\nu=-{6\over{\cal L}}\delta^\mu_\nu \eqno(2.9)$$
which is nothing but the Israel junction conditions~\cite{Isr66} applied to the problem at hand.

When $\alpha\neq0$,  the product of the Dirac and the sign distribution squared is not straightforwardly defined. Indeed, with ${\cal S}^2=1$ in a distributional sense and
${1\over2}{\cal S}'=\delta$, the question is to know whether $\delta{\cal S}^2={1\over2}{\cal S}'{\cal S}^2$ is equal to $\delta$ or, using the Leibniz rule, to ${1\over3}\delta$,
and various proposals have been put forward to give a meaning to (2.8), see [3-4] [6-8] and ~\cite{DerMad03} for a review. They all boil down to obtaining
$$T^w_w=T^w_\mu=0\quad,\quad T^\mu_\nu= -{6\over{\cal L}}\delta^\mu_\nu\left(1-C{4\alpha\over{\cal L}^2}\right)\eqno(2.10)$$
with either $C=1$ (see, e.g, [7]), $C={1\over3}$ (see, e.g., [6]) or $C$ a constant which, it is argued, may depend on the microphysics of the brane, see [8].

When the brane is flat, the difference between the various proposals is immaterial as it amounts to different normalisations of the brane tension. But it matters when one treats
cosmological models for example. Indeed (see the review in~\cite{DerMad03} for details)  one gets for the tension plus matter energy density $\rho\equiv -T^0_0$ the following,
very different, results, depending on whether one has chosen
$C=1$ or
$C={1\over3}$ in (2.10)~:

$$\rho= 6\left(1-{4\alpha\over{\cal L}^2}\right)\sqrt{h^2+{\kappa\over a^2}+{1\over{\cal L}^2}}\qquad\hbox{if}\qquad C=1\eqno(2.11)$$
or
$$\rho= 6\left[1-{4\alpha\over{\cal L}^2}+{8\alpha\over3}
\left(h^2+{\kappa\over a^2}+{1\over{\cal L}^2}\right)\right]\sqrt{h^2+{\kappa\over a^2}+{1\over{\cal L}^2}}\qquad\hbox{if}\qquad C={1\over3}\eqno(2.12)$$
 where  $a(t)$ is the scale factor of the Friedmann-Lema\^\i tre  brane, where $\kappa=+1,0,-1$ characterizes its spatial curvature and  $h\equiv{\dot a\over a}$ is its Hubble
parameter.

Now, from considerations on the proper boundary terms to be added to the action yielding the field equations (2.1), Davis and Gravanis-Willison~\cite{Dav02}
gave a general expression of the stress-energy tensor of matter plus  tension on the brane, in terms of its extrinsic curvature in $V_+$ and its intrinsic Riemann tensor. 

More precisely these authors  associate to a braneworld the following action
$$S=\int_{V_5}\!d^5x\sqrt{-g}\,(-2\Lambda+s+\alpha L_{(2)})+2\!\int_{\Sigma_4}\!d^4x\sqrt{-\bar g}\,{\cal L}_m-2\int_{\Sigma_4}\!d^4x\sqrt{-\bar g}\,Q\,.\eqno(2.13)$$
$g$ is the determinant of the  bulk metric coefficients $g_{AB}$, $\bar g$ that of the induced brane  metric coefficients $\bar g_{\mu\nu}$.  In the second term, ${\cal
L}_m(\bar g_{\mu\nu})$ is the brane ``tension plus matter" Lagrangian. The third, boundary, term is~\cite{Dav02}
$$Q\equiv2K+4\alpha(J-2\bar\sigma^\mu_\nu \,K^\nu_\mu)\eqno(2.14)$$
 with $J$  the trace of
$$ J^\mu_\nu\equiv -{2\over3}K^\mu_\rho K^\rho_\sigma K^\sigma_\nu+{2\over3}K\,K^\mu_\rho K^\rho_\nu+{1\over3}K^\mu_\nu(K.K-K^2)\,.\eqno (2.15)$$
$\bar\sigma^\mu_\nu\equiv\bar r^\mu_\nu-{1\over2}\delta^\mu_\nu\bar s$ is the intrinsic Einstein tensor of the brane $\Sigma_4$ and $K^\mu_\nu$ its  extrinsic
curvature in $V_+$, all indices $\mu$ being moved with $\bar g_{\mu\nu}$ and its inverse $\bar g^{\mu\nu}$.

Thanks to this boundary term the variation of $S$ with respect to the metric coefficients is given in terms of their variations only as~\cite{Dav02}~:
$$\delta S=\int_{V_5}\!d^5x\sqrt{-g}\,\sigma^{[2]}_{AB}\,\delta g^{AB}+\int_{\Sigma_4}\!d^4x\sqrt{-\bar g}\,(2B_{\mu\nu}-T_{\mu\nu})\,\delta\bar g^{\mu\nu}\,.\eqno(2.16)$$
The ``braneworld equations of motion" are therefore  $\delta S=0$, with the metric fixed at the boundaries at infinity only. They are, first,
 the Einstein Gauss-Bonnet ``bulk" equations (2.1) and, second, the brane equations, which generalize the Israel junction
conditions to Einstein Gauss-Bonnet gravity~:
$$B^\mu_\nu\equiv K^\mu_\nu-K\delta^\mu_\nu+4\alpha\left({3\over2}J^\mu_\nu-{1\over2}J\delta^\mu_\nu-
\bar P^\mu_{\ \rho\nu\sigma}K^{\rho\sigma}\right)={1\over2}T^\mu_\nu\eqno(2.17)$$ where
$$\bar P_{\mu \rho \nu \sigma }\equiv \bar R_{\mu \rho \nu \sigma } +(\bar r_{\mu \sigma }\bar g_{\rho\nu}-\bar r_{\rho \sigma }\bar g_{\mu \nu }+\bar r_{\rho\nu}
\bar g_{\mu\sigma }-\bar r_{\mu\nu}\bar g_{\rho\sigma })-{1\over2}\bar s\,(\bar g_{\mu\sigma}\bar g_{\rho\nu}-\bar g_{\mu \nu}\bar g_{\rho
\sigma})\eqno(2.18)$$
 and  where $T_{\mu\nu}$ is defined by $\delta(\sqrt{-\bar g}{\cal
L}_m)\equiv-{1\over2}\sqrt{-\bar g}\,T_{\mu\nu}\,\delta\bar g^{\mu\nu}$ and is interpreted as the stress-energy tensor of ``tension plus matter" on the brane.

When
applied  to a Friedmann-Lema\^\i tre (or flat) brane the brane equations (2.17) reduce to the ``$C=1/3$" result (2.12).

The purpose of this contribution is to confirm this ``$C=1/3$" result using the field equations and the definition of the stress-energy tensor as equal to the ``total brane bending" (eq. 2.7). 

\bigskip
\noindent
{\csc III. A smooth flat brane toy model}
\medskip

Consider a five dimensional spacetime obeying the Einstein Gauss-Bonnet equations with matter, that is such that
$$\sigma^A_{[2]B}={\cal T}^A_B\eqno(3.1)$$
where the Einstein Lanczos tensor is defined in (2.1-2) and where we consider matter to be a scalar field $\phi(x^A)$ with potential $V(\phi)$ and stress energy tensor
$${\cal T}_{AB}=\partial_A\phi\partial_B\phi-g_{AB}\left({1\over2}\partial_C\phi\partial^C\phi+V(\phi)\right)\,.\eqno(3.2)$$

We look for a solution which eventually describes a flat brane embedded in an anti-de Sitter bulk. Hence we consider the metric and scalar field ansatze
$$ds^2|_5=dw^2+g(w)\eta_{\mu\nu}dx^\mu dx^\nu\quad,\quad\phi=\phi(w)\,.\eqno(3.3)$$
It is then a straightforward exercise to find that the field equations
(3.1-2) reduce to
$${\cal T}^w_w=\sigma_{[2]w}^w\qquad\hbox{with}\qquad
{\cal T}^w_w={1\over2}{\phi'}^2-v\quad\hbox{and}\quad 
\sigma_{[2]w}^w={\cal O}\eqno(3.4)$$
$${\cal T}^\mu_\nu=\sigma_{[2]\nu}^\mu\qquad\hbox{with}\qquad
{\cal T}^\mu_\nu=-\left[{1\over2}{\phi'}^2+v\right]\delta^\mu_\nu
\quad \hbox{and}\quad 
\sigma_{[2]\nu}^\mu=({\cal O}+{\cal LB}')\delta^\mu_\nu \eqno(3.5)$$
where
$${\cal O}\equiv -3(k^2-1)(\bar\alpha k^2+\bar\alpha-2)\eqno(3.6)$$
and
$${\cal LB}\equiv k(\bar\alpha k^2-3)\,.\eqno(3.7)$$
A prime denotes a derivative with respect to $z\equiv w/{\cal L}$, ${\cal L}$ being defined by (2.4) ;  $v\equiv{\cal L}^2V$,  $k\equiv -{1\over2}{g'\over g}$
($K^\mu_\nu={k\over{\cal L}}\delta^\mu_\nu$ is the extrinsic curvature of the surfaces
$w=Const.$), and we have introduced the notation
$$\bar\alpha\equiv{4\alpha\over{\cal L}^2}\,.\eqno(3.8)$$
(In accordance with the general properties of the Lanczos tensor the Klein-Gordon equation for $\phi$ is included in (3.4-7), and $\sigma_{[2]w}^w={\cal O}$ is zeroth order in $k'$.
We have gathered in
${\cal LB}'$ all the
$k'$ terms appearing in
$\sigma_{[2]\nu}^\mu$.)

The model must describe a ``smooth brane" in an asymptotically AdS$_5$ bulk with characteristic length scale ${\cal L}$. The following requirements must therefore be met. First
the bulk stress energy tensor ${\cal T}^A_B$ must tend quickly to zero and $k$ to $\pm1$ as $z\to\pm\infty$, so that the metric (3.3) is
asymptotically  AdS$_5$~; second (and this is crucial) $k$ can vary quickly near $z=0$ but must not blow up or behave in such a way that $g$ and hence the metric
 become discontinuous in the thin brane limit.

 Now, there exists, for $0\leq\bar\alpha\leq1$, a very simple toy solution of the field equations (3.4-7), satisfying all these requirements, given by, $A$ being a constant
$$k=\tanh Az\qquad\qquad \left(\Longrightarrow\qquad g={1\over(2\cosh
Az)^{2/A}}\right)\eqno(3.9)$$
which yields
$${\cal O}=-{3\over\cosh^4Az}[2(1-\bar\alpha)\cosh^2Az+\bar\alpha]\eqno(3.10)$$
$${\cal LB}'=-{3A\over\cosh^2Az}[(1-\bar\alpha)\cosh^2Az+\bar\alpha]\eqno(3.11)$$
$${\cal LB}=\tanh Az(\bar\alpha\tanh^2Az-3)\eqno(3.12)$$
as well as
$$v={3\over2\cosh^4Az}[(A+4)(1-\bar\alpha)\cosh^2Az+\bar\alpha(A+2)]\eqno(3.13)$$
$${1\over2}{\phi'}^2={3A\over2\cosh^4Az}[(1-\bar\alpha)\cosh^2Az+\bar\alpha]\,.\eqno(3.14)$$
In the case $\bar\alpha=0$ (Einstein's theory), and in the ``critical" case $\bar\alpha=1$ one obtains $v(\phi)$ in closed form as
$$v(\phi)={3\over2}(A+4)\cos^2\sqrt{A\over3}\phi\qquad\hbox{for}\qquad\bar\alpha=0\eqno(3.15)$$
$$v(\phi)={(A+2)\over6}(A\phi^2-3)^2\qquad\hbox{for}\qquad\bar\alpha=1\,.\eqno(3.16)$$

Let us now look at the thin shell limit, that is the $A\to\infty$ limit,  of this perfectly smooth solution.

First, from (3.9), $g\to e^{-2|z|}$ and hence the metric tends to its bulk AdS$_5$ form everywhere.

Second, from (3.10) (3.13-14), ${\cal O}={1\over2}{\phi'}^2-v$ tends to zero everywhere, but at $z=0$ where it remains finite. We have therefore from (3.4) that $\sigma_{[2]w}^w\sim{\cal T}^w_w\sim0$ in the thin brane limit. More precisely
$$\lim_{A\to\infty}\int_I\sigma_{[2]w}^w\, dz=\lim_{A\to\infty}\int_I{\cal T}^w_w\,dz
=0$$
where $I$ is an interval centered on $z=0$ which eventually goes to zero.

Third, using the following (equivalent) definitions of the Dirac distribution\footnote{Here $f(z)\approx g(z)$ means $\int_If(z)dz=\int_Ig(z)dz$.} 
$$\delta(z)\approx\lim_{A\to\infty}{A\over2\cosh^2Az}\approx\lim_{A\to\infty}{3A\over4\cosh^4Az}\eqno(3.17)$$
we have, from (3.11) (3.13-14), ${\cal O}+{\cal LB}'=-\left({1\over2}{\phi'}^2+v\right)\sim {\cal LB}'\to2(\bar\alpha-3)\delta(z)$, and therefore, from (3.5)
$$\lim_{A\to\infty}\sigma_{[2]\nu}^\mu\approx2(\bar\alpha-3)\delta^\mu_\nu\,\delta(z)\approx\lim_{A\to\infty}{\cal T}^\mu_\nu\,.\eqno(3.18)$$
Comparing this equation with (2.7) (and recalling that $\delta(w)={\cal L}\delta(z)$) we see that we are led to identify
$$T^\mu_\nu\equiv{2\over{\cal L}}(\bar\alpha-3)\delta^\mu_\nu=-{6\over{\cal L}}\left(1-{4\alpha\over{3\cal
L}^2}\right)\delta^\mu_\nu\eqno(3.19)$$
to the brane stress-energy tensor (or, rather, brane tension in that
case).

 On this simple toy model we hence recover the ``C=1/3'' result advocated in [4] [6] [9].\footnote{Alternative definitions for the brane stress-energy tensor can however be put
forward. In~\cite{Ger03} the brane stress-energy tensor is defined as the bulk stress-energy tensor ${\cal T}^A_B$ evaluated at the particular point
$z_s$ ($s$  for ``screen"), such that $k'(z_s)=2/{\cal L}$. With such a definition one gets $T^\mu_\nu=-{6\over{\cal L}}\left(1-{4\alpha\over{\cal
L}^2}\right)\delta^\mu_\nu$, that is the ``C=1" result.This ``screen" hypersurface $z=z_s$ ``stores" the information of the total bending of the brane and can be defined for any smooth function $k'(z)$ which tends to a Dirac distribution $\delta(z)$, hence rendering the result general. Moreover since $\lim_{A\to\infty}z_s=0$ the screen is inside the domain wall. See~\cite{Ger03}for details.}

Let us conclude this section with a remark which will be useful in section V.  From (3.12) one notes, that
$$\lim_{A\to\infty}{\cal LB}=(\bar\alpha-3){\cal S}(z)\eqno(3.20)$$
where ${\cal S}(z)$ is the sign distribution such that ${\cal S}'=2\delta$. Therefore
the brane stress-energy tensor (3.19) is also given by
$$T_\nu^\mu=2B\delta_\nu^\mu\eqno(3.21)$$
where ${\cal LB}'\delta^\mu_\nu$, which contains all the $k'$-terms, is the dominant
part of the Lanczos tensor when $A\to\infty$ and where $B$ is the AdS$_5$ value of ${\cal B}$ (3.7) evaluated at $w=0_+$, that is with $k=+1$.

 Of course, it remains to show that the result (3.19) is not model dependent, that is, does not depend on the particular choice made in (3.9) for $k(z)$ (or, equivalently, on the
particular choice (3.13-16) for $v(\phi)$), it being understood though that the requirements listed above remain satisfied.\footnote{Indeed, consider for example the other
ansatz~:
$k=\tanh Az\left(1+{\beta\over\cosh^2Az}\right)$ which yields the metric $\ln g=-\ln [(2\cosh
Az)^{2/A}]+{\beta\over A\cosh^2Az}$. If $\beta$ remains finite when $A\to\infty$ then $\ln g\sim -2|z|\ \forall z$ and this ansatz, as can be easily seen,  yields the same
brane tension as the
$\beta=0$ case treated in the text. If $\beta=\bar\beta A^n$ with $n>0$ then the brane stress energy tensor is no longer given by (3.19). However such ansatze must be
disgarded as the metric $\ln g$ is then no longer continuous (for example, for $n=2$, $\ln g\sim -2|z|+2\bar\beta\delta(z)$).}

\bigskip
\noindent
{\csc IV. Model independence of the thin flat brane tension}
\medskip

 Consider an arbitrary confining potential $v(\phi)$ that is such that 
$$v[\phi(z)]=v_0\delta_A(z)\eqno(4.1)$$
 where $\delta_A(z)$ is any function which tends to a distribution localized at $z=0$ when the parameter $A\to\infty$.

 We look for a solution of the field equations (3.4-7) such that $k$
 is everywhere finite when $A\to\infty$ and  tends to $\pm1$ when
 $z\to\pm\infty$, and such that $k'$, like $v$, is ``confining",
 i.e. picked on $w=0$. Hence, for large $A$~: ${\cal O}\ll {\cal LB}'$.

 Therefore equation (3.4-5) yield, for large $A$,
$${1\over2}{\phi'}^2\sim v\qquad,\qquad {\cal LB}'=[k(\bar\alpha k^2-3)]'\sim-2v\,.\eqno(4.2)$$
Since $k$ is everywhere finite, ${\cal LB}$ is also everywhere finite and hence ${\cal LB}'$ cannot do else than tend to the Dirac distribution. This implies that $\delta_A(z)$ must be 
 such that $\int_{-\infty}^{+\infty}\delta_A(z)=1$. Integrating we then get
$$\left[k\left(\bar\alpha k^2-3\right)\right]_{-\infty}^{+\infty}\sim-2\int_{-\infty}^{+\infty}dz\,v(z)=-2v_0\,.\eqno(4.3)$$
Now, $k(\pm\infty)=\pm1$. Hence
$$v_0=3-\bar\alpha\,.\eqno(4.4)$$
This result just means that the potential must be ``fine-tuned" in order not to introduce a extra, spurious, cosmological constant in the model.

Returning to (3.5) we hence have
$$\sigma_{[2]\mu}^\nu\sim
{\cal LB}'\delta^\mu_\nu\sim2(\bar\alpha-3)\delta^\mu_\nu\delta_A(z)\to2(\bar\alpha-3)\delta^\mu_\nu\delta(z) \eqno(4.5)$$
and therefore, from the definition (2.7)
$$T^\mu_\nu={2\over{\cal L}}(\bar\alpha-3)\delta^\mu_\nu=-{6\over{\cal L}}\left(1-{4\alpha\over3{\cal L}^2}\right)\delta^\mu_\nu\,.\eqno(4.6)$$
Hence we see that the expression for the brane stress-energy tensor
obtained in the previous section does not depend on the  specific form chosen for the confining potential $v(\phi)$.

As for the bulk stress-energy tensor it is, still for large $A$
$${\cal T}^\mu_\nu\sim-2v\,\delta^\mu_\nu\sim-2\delta^\mu_\nu(3-\bar\alpha)\delta_A(z)
\quad\hbox{so that}\quad\lim_{A\to\infty}{\cal
T}^\mu_\nu\approx-{6\over{\cal L}}\left(1-{4\alpha\over3{\cal
L}^2}\right)\delta^\mu_\nu\,\delta(w)\,.\eqno(4.7)$$
 Hence we check that the expression for the bulk stress-energy tensor
obtained in the previous section was not model dependent either.

Let us also, for completeness, give the expressions for the bulk metric and scalar field (at leading order in $A$). 

From (4.1-4) we have that $k$ is the (unique) solution which tends to $1$ at $z\to+\infty$ of the cubic equation
$$k(\bar\alpha k^2-3)\sim-(3-\bar\alpha)S_A(z)\eqno(4.8)$$
where $S_A'(z)=2\delta_A(z)$ is such that $S_A(+\infty)=1$. In the
limit $A\to\infty$ $S_A$ tends to the sign distribution ${\cal S}$ and 
$$\lim_{A\to\infty}k={\cal S}\eqno(4.9)$$
and therefore the metric reduces to (2.5).

Finally, from (4.1-2) (4.4) we have that
$$\phi(z)\sim\sqrt{2(3-\bar\alpha)}\int\sqrt{\delta_A(z)}dz\eqno(4.10)$$
and, hence, $v(\phi)=(3-\bar\alpha)\delta_A(z)$ is known as a function of $\phi$, at least
implicitely. It is clear that to different functions $\delta_A(z)$
(two examples being displayed in eq (3.15)) will correspond  different
$v(\phi)$ (e.g. (3.13) or (3.14)). However, whatever the value of
$\bar\alpha$, these potentials yield the thin brane limit (4.6), (4.7)
when $A\to\infty$ (this ``loss of information'' being the reason why
the brane stress-energy tensor becomes model independent in the thin
brane limit).

\bigskip
\noindent
{\csc V. Generalization to curved branes}
\medskip

In a Gaussian normal coordinate system adapted to some timelike foliation
$$ds^2|_5=dw^2+\gamma_{\mu\nu}(w,x^\rho)dx^\mu dx^\nu\eqno(5.1)$$
where
$${\cal K}_{\mu\nu}=-{1\over2}{\partial\gamma_{\mu\nu}\over\partial w}\eqno(5.2)$$
is the extrinsic curvature of the surface $w=Const.$, the Einstein Gauss-Bonnet equations
$$\sigma_{[2]B}^A={\cal T}^A_B\eqno(5.3)$$
 split into
$${\cal T}^w_w=\sigma_{[2]w}^w\quad\qquad\hbox{with}\quad\qquad \sigma_{[2]w}^w={\cal O}\eqno(5.4)$$
$${\cal T}^\mu_\nu=\sigma_{[2]\nu}^\mu\quad\qquad\hbox{with}\quad\qquad \sigma_{[2]\nu}^\mu={\cal O}^\mu_\nu+
{\partial{\cal B}^\mu_\nu\over\partial w}\eqno(5.5)$$
where ${\cal O}$ and ${\cal O}^\mu_\nu$ are quartic in the extrinsic curvature and where ${\cal B}^\mu_\nu$ is given by~\cite{DerDoz00}
$${\cal B}^\mu_\nu= {\cal K}^\mu_\nu-{\cal K}\delta^\mu_\nu+4\alpha\left({3\over2}{\cal J}^\mu_\nu-{1\over2}{\cal J}\delta^\mu_\nu-
\bar {\cal P}^\mu_{\ \rho\nu\sigma}{\cal K}^{\rho\sigma}\right)\eqno(5.6))$$ with
$$ {\cal J}^\mu_\nu=-{2\over3}{\cal K}^\mu_\rho {\cal K}^\rho_\sigma {\cal K}^\sigma_\nu+{2\over3}{\cal K}\,{\cal K}^\mu_\rho {\cal K}^\rho_\nu+{1\over3}{\cal K}^\mu_\nu({\cal K}.{\cal K}-{\cal K}^2)\,.\eqno (5.7))$$ and
$$ {\cal P}_{\mu \rho \nu \sigma }={\cal R}_{\mu \rho \nu \sigma } +({\cal R}_{\mu \sigma }\gamma_{\rho\nu}-{\cal R}_{\rho \sigma }\gamma_{\mu \nu }+{\cal R}_{\rho\nu}
\gamma_{\mu\sigma }-{\cal R}_{\mu\nu}\gamma_{\rho\sigma })-{1\over2}{\cal R}\,(\gamma_{\mu\sigma}\gamma_{\rho\nu}-\gamma_{\mu \nu}\gamma_{\rho
\sigma})\eqno(5.8)$$
where ${\cal R}_{\mu\nu\rho\sigma}$, ${\cal R}_{\mu\nu}$ and ${\cal R}$ are the Riemann tensor, Ricci tensor and scalar curvature of the surface $w=Const.$.
(The fact that all the terms containing a $w$-derivative of the extrinsic curvature can be gathered in a $w-$ derivative of a tensor ${\cal B}^\mu_\nu$ is not trivial and is particular to the Lanczos tensor.)

 If all matter is to be confined on the surface $w=0$ and the metric remain continous then ${\cal O}$, ${\cal O}^\mu_\nu$ and ${\cal B}^\mu_\nu$ will remain finite, while ${\partial{\cal B}^\mu_\nu\over\partial w}$ will tend to a delta distribution localized at $w=0$.
 More precisely, if the bulk is imposed to be almost anti-de Sitter for all $w$ larger than, say, $w_0>0$, with $w_0\to0_+$, that is if
$${\cal K}^\mu_\nu\sim{1\over{\cal L}}\eta_{\mu\nu}\qquad\forall w>w_0,\qquad w_0\to0_+\eqno(5.9)$$(with ${\cal L}$ given by (2.4)), then ${\cal O}_w\sim{\cal O}^\mu_\nu\sim0$ and
$${\cal B}^\mu_\nu\sim B^\mu_\nu{\cal S}(w)\eqno(5.10)$$
with $B^\mu_\nu={\cal B}^\mu_\nu(0+)$ and ${\cal S}$ the sign distribution. Consequently
$$\sigma_{[2]\nu}^\mu\sim {\partial{\cal B}^\mu_\nu\over\partial w}\to 2B^\mu_\nu\delta(w)\,.\eqno(5.11)$$  
Hence, from the definition (2.7)
$$T^\mu_\nu=2B^\mu_\nu\eqno(5.12)$$
which generalizes (3.21),
is to be identified with the brane matter plus tension stress energy tensor. Therefore, in the general case as well, the brane equations (2.17) obtained by Davis and Gravanis-Willison are recovered.

\bigskip

{\bf Acknowledgements~:} 
N.D. thanks Joseph Katz for fruitful  discussions about matching conditions in gravity theories, and the Yukawa Institute for Theoretical
Physics where this work was done,  for its hospitality.C.G. thanks Carlos Barcelo and Carlos F. Sopuerta for useful discussions.

\bigskip

\providecommand{\href}[2]{#2}\begingroup\raggedright\endgroup

\end{document}